\documentclass[aps,10pt,prl,twocolumn,superscriptaddress,showpacs,floats,floatfix,preprintnumbers,amsmath,amssymb]{revtex4-1}
\usepackage[T1]{fontenc}
\usepackage[latin9]{inputenc}
\usepackage[USenglish,american]{babel}
\usepackage{amsmath}
\usepackage{amssymb}
\usepackage{wasysym}
\usepackage{graphicx}
\usepackage{tabularx}
\usepackage{xcolor}
\usepackage{bm}          
\usepackage{pifont} 
\usepackage{dcolumn}
\usepackage{float}
\usepackage{natbib}
\usepackage{enumerate}

\usepackage{subfigure,textcomp,epsfig,siunitx}
\usepackage{array}
\usepackage{longtable}
\usepackage{bm}
\usepackage{booktabs}

\graphicspath{{plots/}}

\newcommand{\GrazPhys}{Institute of Theoretical and Computational Physics,
Graz University of Technology, NAWI Graz, 8010 Graz, Austria}
\newcommand{\UniRoma}{Dipartimento di Fisica, Universit\`a di Roma La Sapienza, Piazzale Aldo Moro 5, I-00185 Roma, Italy}

\begin{document}
\title{Fused borophenes: a new family of superhard materials}

\author{Santanu Saha} \affiliation{\GrazPhys}
\email{santanu.saha@tugraz.at}
\author{Wolfgang von der Linden} \affiliation{\GrazPhys}
\author{Lilia Boeri} \affiliation{\UniRoma}

\date{\today}

\begin{abstract}
The search of new superhard materials has received a strong impulse by industrial demands for low-cost 
alternatives to diamond and $c$-BN, such as metal borides. In this Letter we introduce a new family of
superhard materials, "{\it fused borophenes}", containing  2D boron layers which are interlinked to
form a 3D network.
These materials, identified through a high-throughput scan of B$_x$C$_{1-x}$ structures, exhibit Vicker's 
hardnesses comparable to those of the best commercial metal borides.
Due to their low formation enthalpies, fused borophenes could be synthesized by high-temperature methods, 
starting from appropriate precursors, or through quenching of high-pressure phases.
\end{abstract}

\maketitle
\textbf{Introduction}-
The revamped interest in superhard materials has been driven not only by scientific curiosity, but also by
the increasing technological interest, in several industrial applications
~\cite{solozhenko2019hunt,kvashnin2019computational}.
Diamond, with a reported Vicker's Hardness ($V_H$) of 120 GPa, holds by far the record among all known materials, 
but its chemical reactivity at high temperatures and the high production cost restrains its practical usability. 
Improvements over the current alternatives, such as cubic-BN ($c$-BN) or 
cubic-BC$_2$N~\cite{wakatsuki1972synthesis,zhang2006structural,solozhenko2001synthesis}
and metal borides~\cite{gu2008transition,levine2009advancements,liang2013thermodynamic},
which also present serious limiting issues such as high synthesis price, or limited hardness, are being intensively sought.

The phase diagram of elements such as boron, carbon, nitrogen and oxygen and their compounds represents an ideal
hunting ground for potential superhard materials to be explored by {\em ab-initio} methods for crystal structure
prediction(CSP) and high-throughput (HT) screening, which are rapidly expanding the scope of material research~\cite{curtarolo2013high,flores2020perspective,kvashnin2019computational,BCN:ML,avery2019predicting,fujii2020pentadiamond,saha2020comment,brazhkin2020comment}.
 
%

%
%

In this paper, applying a combination of crystal structure prediction (minima hopping)~\cite{goedecker2004minima,amsler2010crystal} 
and high-throughput screening techniques to the boron-carbon (B-C) phase diagram, we uncover a new family of
metastable boron and boron-rich carbon structures, "{\it fused borophenes}" (FB) with hardness and elastic properties 
comparable to those of the best  metal borides.
\footnote{
{\bf Computational Methodology-} 
The plane-wave based Vienna Ab-initio Simulation Package (VASP)~\cite{VASP_Kresse} DFT package has been used for 
geometric relaxation; and calculation of total energies, electronic properties, Electron Localization 
Function(ELF) and elastic properties. The calculation of phonon dispersion has been carried out using Quantum-Espresso-6.4.1~\cite{giannozzi2017advanced,baroni2001phonons}. Unless mentioned, all the DFT calculations 
have been carried out using Local Density Approximation (LDA) exchange-correlation functional. The in-built LDA 
Projector Augmented Wave potentials have been used for VASP~\cite{kresse1999ultrasoft} calculations, whereas ONCV 
norm-conserving potentials~\cite{hamann2013optimized} have been used for QE. Details of the calculations can be 
found in the Supplementary Material(SM).}
Fused Borophenes, as the term suggests, can be seen as a stacking of different types of 2D boron layers connected
through covalent bonds to form 3D bulk structures. Being structurally related to the high-pressure $\alpha$-Ga 
phase, considered to be the most likely candidate to explain superconductivity observed in boron above 160 GPa
~\cite{eremets2001superconductivity,papaconstantopoulos2002first,ma2004electron}, FBs ideally represent the 
missing link between the two known families of boron structures: 2D boron mono- and bilayers (borophenes)
~\cite{sun2017two,wang2019review,gao2018structure}, and 3D bulk structures based on B$_{12}$ icosahedral units, 
such as $\alpha$, $\beta$ and $\gamma$ boron, ~\cite{oganov2009boron,penev2012polymorphism,wang2019review}  
FBs are metastable at ambient conditions, but our calculations suggest that quenching from high pressure may be
used to stabilize some of the most competitive phases.

In the following, after discussing the general elastic and thermodynamic properties of superhard fused borophenes, 
we will introduce a classification scheme into three families, based on the structural motifs; after this, we will 
discuss the origin and nature of the exceptional elastic properties, with a detailed analysis of the three hardest
structures in each family ($\alpha^*$, $\beta^*$ and $\gamma^*$ FBs).
\begin{figure*}[!htb]
\centering
\includegraphics[width=1.6\columnwidth,angle=0]{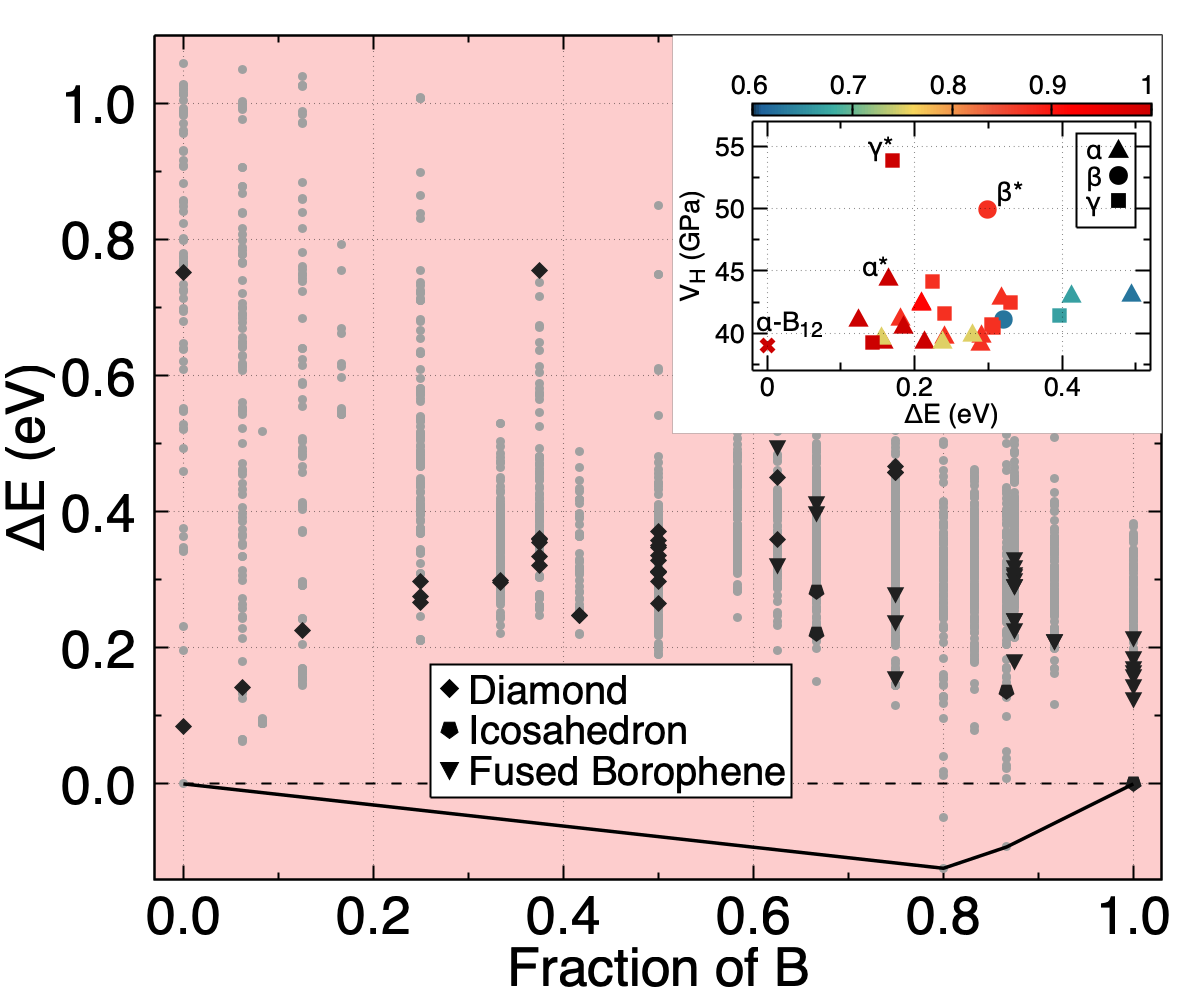}
\caption {Relative energy $\Delta$E(eV/atom) vs fraction of B of all the predicted $\sim$2700 B-C structures.
Superhard diamond-like(diamond), icosahedron-like (invered pentagon) and fused borophenes FB (inverted triangles) 
are shown by black coloured symbols. Rest are shown by grey circles.
(Inset): Relative energy $\Delta$E(eV/atom) vs Vicker's hardness V$_H$(GPa)
of superhard FB. V$_H$ was calculated using Chen's formula.
The 17 $\alpha$, 2 $\beta$ and 8 $\gamma$ superhard FB are shown with triangles,
circles and squares, respectively. The three hardest fused borophenes 
$\alpha$*, $\beta$* and $\gamma$* are marked in the plot. For reference, 
we also indicated the position of $\alpha$-B$_{12}$ (red cross). The colour represent 
the fraction of B in the FB. 
The relative energy in both the main and inset figure is calculated with respect to
C-graphite and icosahedral  $\alpha$-B$_{12}$ boron.}
\label{fig:fig1}
\end{figure*}

\textbf{Fused Borophenes in the B-C Phase Diagram-}
Fused borophenes were identified through a HT search of hard materials on a database of $\sim$2700 distinct 
B-C structures, obtained through an unbiased minima hopping search~\cite{goedecker2004minima,amsler2010crystal}, 
on 8-15 atoms unit cells, with variable B/C composition.~\cite{saha2020high} Fig.\ref{fig:fig1} shows the location
of all resulting structures in formation energy ($\Delta E$) vs Boron fraction plane;
C-graphite and $\alpha$-B$_{12}$ were used to compute the reference energy.

Through HT screening of the initial set, we singled out 71 B-C structures with Vicker Hardness 
(V$_H$) $\geqslant$ 40 GPa, which is the conventional threshold for superhard materials: these are shown as
black symbols in Fig.\ref{fig:fig1}.
Among these 72 superhard materials, 40 are diamond-like  carbon-rich structures (filled diamonds). Based on the
atomic arrangement, the remaining 31 boron-rich compounds can be classified in two groups: 4 are icosahedron-like 
(filled inverted pentagons), while 27 of them can neither be classified as $B_{12}$ icosahedra nor 
triangular/hexagonal  2D boron sheets (borophenes). We termed this new class of boron-rich B-C structures as 
"{\it fused borophenes}" (FB),  and indicate them in the plot as black inverted triangles.
FBs appear in the B-C phase diagram at a minimum boron concentration  
B$_{7/12}$C$_{5/12}$, with  relative energies $\Delta$E $\geqslant$ 293 meV/atom
and become gradually more stable as the B fraction increases.
There is a total of 306 FB structures in the whole database of 2700 B-C structures.
Among them only 27 have V$_H$ $\geqslant$ 40 GPa, i.e. $\sim$9 \% of FB structures are superhard materials.

The inset of Fig.\ref{fig:fig1} displays a plot of the Vicker's Hardness V$_H$ (GPa) against relative energy
of the 27 FB superhard structures.
The merged plot of the bulk modulus B (GPa) along with the Vicker's Hardness V$_H$ (GPa) against relative energy
of the 27 FB superhard structures is provided in the supplementary material(SM). The colour of the symbols indicates
the boron fraction in the structure, from pure boron (red) to the minimum B concentration (blue). Depending on the
nature and relative arrangement of the boron layers, FB have been divided into three families: $\alpha$ (triangles), 
$\beta$ (circles) and $\gamma$ (squares).
%
%
%
There is no evident correlation between the formation energy $\Delta$E and
the V$_H$ of the FB in the inset of Fig.\ref{fig:fig1}.

It is encouraging to observe that all superhard FBs shown in the plot have formation energies $\Delta$E $\leqslant$ 
500 meV/atom, which implies that they could be synthesized under appropriate conditions
~\cite{aykol2018thermodynamic,saha2020high}.
For a reference, among the different borophene phases synthesized till date on Ag/Au substrates, the lowest-energy
borophene is 2-Pmmn phase with 555 meV/atom w.r.t. $\alpha$-B$_{12}$~\cite{lherbier2016electronic}.

As for the hardness, most structures have a V$_H$ close to the threshold value of 40 GPa.
A few structures, however, stand out as exceptional, with $V_H$ exceeding 45 GPa:
among these, we selected one structure per prototype, indicated as $\alpha^*$, $\beta^*$ and $\gamma^*$
in Fig.\ref{fig:fig1}, which will be discussed in detail in the following. 
%


{\bf Classification of Fused Borophenes in Different Families} - 
Like other known boron polymorphs, Fused Borophenes exhibit a large diversity of motifs, 
reflecting the electron-deficient nature of this element~\cite{boustani1997systematic}.
%
Fused borophenes identified in our search may contain: (i)  boron layers(BL) of different types and
(ii) single or V-shaped double bonds connecting one atom in a layer to two atoms in another layer. 
Furthermore, (iii) a structure may comprise BL all of the same type or of different types, and 
(iv) their stacking along the vertical axis may also be different.
\footnote{All FBs discussed here contain a maximum of two layers; this is
likely due to limit of 15 atoms/unit cell imposed in our initial structural search.}

In order to uniquely identify all structures generated we assigned each FB a unique alphanumeric ID,
of the form: "TYPE-XYZ(HH)-XYZ(HH)". Here, TYPE indicates the family -- $\alpha$, $\beta$ or $\gamma$; 
the XYZ numerals denote the polygonal motif present in each of the two BL, while HH is the hexagonal-hole 
concentration of the BL if applicable. The XYZ(HH) notation is introduced for each BL in the structure; 
the unconventional BL precedes the conventional BL.
The family prototype, $\alpha$, 
$\beta$ or $\gamma$, is assigned on the basis of the type and sequence of stacked boron layers (BL):
\begin{itemize}
    \item {\bf $\alpha$}: FBs formed by stacking {\em conventional} BL only. 
    \item {\bf $\beta$}: FB consisting of both {\em conventional} and {\em unconventional} BL. 
    \item {\bf $\gamma$}: FB made of {\em unconventional} BL only.
\end{itemize}

Boron layers are termed {\em conventional}, if they exhibit only triangular and hexagonal patterns, 
as quasi-2D BLs synthesized in mono- and bilayer form (borophenes);
~\cite{tang2007novel,mannix2015synthesis,feng2016experimental,gao2018structure,yao2019observation} 
or {\em unconventional} if they contain other polygonal motifs.
Within this classification scheme, the $\alpha$-Ga structure,
formed by stacking two triangular BLs with 0 HH concentration
would be assigned the ID $\alpha$-3(0)-3(0).

\begin{figure}[!htb]
\centering
\includegraphics[width=1.0\columnwidth,angle=0]{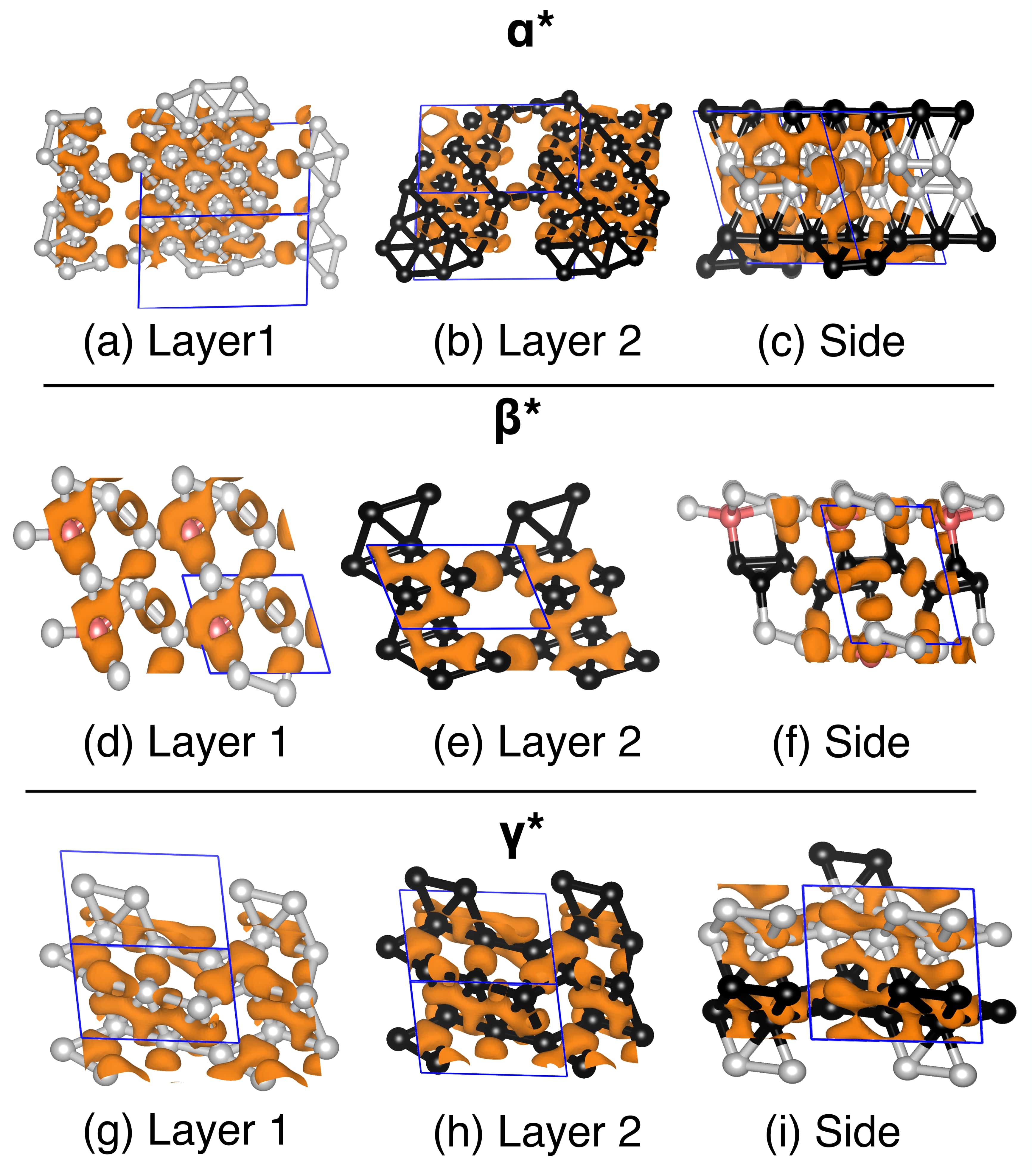}
\caption {The top and side view of BL of the three best superhard FB
materials: 
$\alpha$*, 
$\beta$*, 
and $\gamma$*. 
The white and black colour sphere indicate B atoms belonging to different BLs.
Red spheres indicate C atoms in layer 1 of $\beta$*. The unit cell is shown as a blue line.
Isosurfaces of the 
Electron localization function (ELF) at $\eta\approx$0.70 of $\alpha$*(a-c), $\beta$*(d-f) and
$\gamma$*(g-i) FBs  are superimposed in orange to the three structures.
}
\label{fig:BestStruc}
\end{figure}

Examples of structures belonging to each of the three families are shown in Fig.\ref{fig:BestStruc}. 
In particular, the figure depicts the three structures indicated as $\alpha^*$, $\beta^*$ and $\gamma^*$.
$\alpha$* FB ($\alpha$-36(2/25)-36(2/25)) - comprises two identical conventional BL with 2/25 HH concentration, 
interconnected through V-shaped double bonds (Fig.\ref{fig:BestStruc}a-c); $\beta$* FB, with composition 
 B$_{7/8}$C$_{1/8}$ ($\beta$-356-36(1/5))
 comprises an unconventional BL (Fig.\ref{fig:BestStruc}(d)) with a triangular-pentagonal-hexagonal motif and a
 conventional BL (Fig.\ref{fig:BestStruc}(e)) with 1/5 HH concentration. These two BL are connected by single bonds as
 shown in Fig.\ref{fig:BestStruc}(f). $\gamma$* FB  ($\gamma$-34-34) comprises two unconventional BL of the same type, 
 interconnected through a V-shaped double bond (Fig.\ref{fig:BestStruc}(g)-(i)). 

In all three structures, bonds within and between the BLs are of comparable length.  The isosurfaces of the 
Electronic Localization Function (ELF) - $\eta\approx$ 0.7 - superimposed to the three structures show that
the electronic charge tends to accumulate both along the bonds connecting  atoms in the same BLs and between
different layers. Thus, FBs should  be described rather as a bulk-like interconnected network of boron layers, 
and not as a collection of weakly-bound layers, quasi-2D van-der-Waals systems.
This peculiar structural arrangement of FBs is at the heart of their remarkable elastic properties.

\begin{table}[!htb]
\centering
\begin{tabular}{lccccccccc}
\hline 
ID   & $\Delta$E & Vol & State & $B$ & $G$ & $Y$ & V$_H$ & $\nu$ & $A^U$ \\ 
\cmidrule{5-8}
   &  (meV) & (\AA$^3$) &  & \multicolumn{4}{c}{(GPa)} &  & \\ 
\hline\hline 
\multicolumn{10}{l}{Fused Borophene} \\
$\alpha$* & 164 & 6.68 &  M & 245 & 236 &  547 & 44 & 0.136 & 0.362 \\ 
$\beta$* & 278 & 6.53 &  M & 242 & 246 &  558 & 48 & 0.121 & 0.190 \\ 
$\gamma$* & 177 & 6.35 & NM & 270 & 279 &  631 & 53 & 0.115 & 0.247 \\ 
\hline 
\multicolumn{10}{l}{Other Covalent Systems} \\
C-Dia  &  085   & 5.52  & NM & 454 & 537 & 1155 & 93 & 0.076 & 0.046 \\ 
$c$-BN   &  -- & 5.75  & NM & 409 & 407 &  916 & 64 & 0.126 & 0.200   \\ 
$\alpha$-B$_{12}$ & 000 & 7.00 & NM & 225 & 210 &  480 & 39 & 0.145 & 0.192 \\ 
B$_6$O &  -- & 7.11 & NM & 239 & 211 &  496 & 37 & 0.158 & 0.239  \\ 
\hline 
\multicolumn{10}{l}{Metal Borides} \\
TiB$_2$ & --  & 8.20 & M & 282 & 282 & 638 & 51 & 0.125 & 0.123 \\
CrB$_4$ & --  & 7.14 & M & 300 & 277 & 647 & 46 & 0.148 & 0.347 \\
ReB$_2$ & --  & 8.89 & M & 370 & 296 & 711 & 40 & 0.184 & 0.193 \\
\hline 
\multicolumn{10}{l}{Quasi-2D vdW System} \\
Graphite & 000 & 8.56 &  & 165 & 119 & 803 & 19 & 0.209 & 93.489 \\
\end{tabular}
\caption{Summary of the calculated properties of the three best superhard FB: $\alpha$*,
$\beta$* and $\gamma$*, marked in Fig.\ref{fig:fig1}. The properties of 
$\alpha$-B$_{12}$, Carbon diamond(C-Dia), and of a few other covalent and metal boride hard materials
are listed for reference. In addition, the properties of quasi-2D materials graphite
is also listed. Starting with the ID of the structures in the first column, the
relative energy ($\Delta$E) w.r.t. C-graphite and $\alpha$-B$_{12}$ in meV/atom and volume(Vol)
in \AA$^3$/atom are listed in 2nd and 3rd column. The metal(M)/non-metal(NM) state is mentioned
in 4th column. The remaining columns contain mechanical properties: bulk modulus ($B$),  shear modulus ($G$),
Young's modulus ($Y$) and Vicker's hardness (V$_H$)in GPa. The Poisson's ratio $\nu$ and the universal
elastic anisotropy index $A^U$ are listed in column 9 and 10, respectively.}
\label{tab:bestdetails}
\end{table}

{\bf Thermodynamical and Mechanical Properties of $\alpha^*$, $\beta^*$ and $\gamma^*$ fused borophenes} -

Table ~\ref{tab:bestdetails} contains selected thermodynamic, geometric and elastic properties 
of  $\alpha$*,  $\beta$*,  and $\gamma$* FBs, 
calculated at the LDA-DFT level
\footnote{
{\bf Computational details of elastic properties calculations}:
The elastic properties of the enlisted systems in Table~\ref{tab:bestdetails} were estimated from
the elastic tensor based on the Voigt-Reuss-Hill approximation~\cite{hill1952elastic}. The Chen's
model~\cite{chen2011modeling} was used to estimate the hardness. 
The elastic tensor was calculated with a two step displacement of
$\delta$=$\pm$0.005~\AA,$\pm$0.01~\AA. The crystal structures of other hard covalent systems 
and metal borides in Table~\ref{tab:bestdetails} were obtained from Materials Project
database~\cite{jain2013commentary}. These structures were further geometrically relaxed and their 
different properties were reevaluated. Details of the calculations can be found in the
Supplementary Material(SM).}.
The table also reports for comparison, the same properties calculated for other well-known superhard covalent
materials and metal borides, as well as for graphite,
as a representative example of quasi-2D van-der-Waals system.

With the exception of diamond and $c$-BN, the calculated V$_H$ of $\alpha$*, $\beta$*, and $\gamma$* FBs is
larger than other superhard materials listed in the table. Their formation energies are relatively low 
($\Delta$ E $\leqslant$ 280 meV/atom). Furthermore, the three phases satisfy the general stability criteria, 
i.e. the elastic tensor $C_{ij}$ is positive definite and has positive eigenvalues, and are dynamically stable. 
While $\alpha$* and $\beta$* FBs are both metallic, $\gamma$* is a semiconductor, with a small gap of 0.46 eV. 
Additional details on the calculations are reported in the SM, together with electronic and phononic spectra.

The three elastic moduli, i.e. the bulk ($B$), shear ($G$) and Young's ($Y$) moduli, are considerably smaller 
than the covalent superhard materials with a $sp^3$ tetragonal arrangement, such as  C-Diamond (C-Dia) and 
$c$-BN, but comparable or larger than the other hard non-tetragonal covalent systems ($\alpha$-B$_{12}$ and 
B$_6$O), and hard metal borides. 
The  Poisson's ratio $\nu$ and universal elastic anisotropy index $A^U$~\cite{ranganathan2008universal} are in line with those of other bulk 
systems, but sensibly different than in graphite, confirming that FBs have substantially a bulk nature. 

Having established that FBs are competitive compared to other classes of widely-used superhard materials, 
two obvious questions remain to be asked: What is the origin of their record hardness? What strategies
can be used to synthesize them?

The reasons underlying the exceptional hardness of superhard materials have been extensively investigated, 
revealing two different mechanisms: diamond and other $sp^3$ materials, such as $c$-BN and $c$-BC$_2$N, 
contain a dense lattice of strong covalent bonds which are hard to compress; in metal borides, on the 
other hard, regions of incompressible electronic densities associated to the large metal ions limit 
the overall compressibility of the materials.~\cite{kaner2005designing,cumberland2005osmium,chung2007synthesis}.
Clearly, it is the first mechanism which is at play in FBs: the atomic volumes are sensibly smaller than in 
metal borides and, although larger than in C-Dia or $c$-BN, absolutely in line with those of other covalent 
borides reported in the table.

Concerning possible strategies to synthesize FBs, a first option is to resort to high-temperature reactions, 
starting from appropriate precursors. The predicted enthalpies of formation of $\alpha$*, $\beta$* and $\gamma$* 
FBs are within typical stabilization energies of metastable polymorphs.~\cite{aykol2018thermodynamic}
For a few cases, an alternative route would be quenching from high pressure. In particular, since,
as shown in Fig.1 of SM, $\gamma$* emerges as a competitive phase due to lowering of enthalpy 
for pressures $\geqslant$ 100 GPa, a sample of boron compressed at these pressures and slowly brought 
to ambient pressure may retain this structure.

\textbf{Conclusions}-
In conclusion, in this paper we presented a new class of superhard materials, {\it fused borophenes}, 
identified through HT screening of B-C structures generated in a minima-hopping run. FB can be seen as a 
stacking of 2D boron-layers, interconnected through either a single or V-shaped double covalent bonds, 
which makes them distinct from other known 2D boron materials (borophenes) reported in literature. 
The spatial distribution of the Electronic Localization Function and the low universal elastic anisotropy 
index $A^U$ clearly show that, due the presence of covalent bonds between the atoms of the interconnected BLs, 
FBs effectively behave as 3D bulk structures, although they are composed of 2D BLs. 

Due to their relatively low formation energies, these systems, which are strong contenders to hard metal 
borides as superhard materials for industrial applicatitons, may be synthesizable through high-temperature 
synthesis, starting from appropriate precursors. Furthermore, our calculations hint that the structure labeled 
as $\gamma^*$ in this work, which, with a predicted $V_H$ of 53 GPa, second only to C-diamond and $c$-BN, may 
be synthesized by quenching from high pressures.

\textbf{Acknowledgements}-
S.Saha and W. von der Linden acknowledge computational 
resources from the dCluster of the Graz University of Technology and the VSC3 of the Vienna 
University of Technology, and support through the FWF, Austrian Science Fund, Project
P 30269- N36 (Superhydra). L. Boeri acknowledges support from Fondo Ateneo Sapienza 
2017-18 and computational Resources from CINECA, proj. Hi-TSEPH.

{\bf Note added:} While completing the present manuscript, a preprint has appeared
~\cite{hilleke2021structural}, which reports layered structures of boron, discovered 
by crystal structure prediction with evolutionary algorithms. 
The authors propose a different classification scheme of fused borophenes
into {\it derivatives of the $\alpha$-Ga } and {\it channel} structures. 
The {\it Channel-I} structure is analogous to our $\gamma^*$ structure.
The main results of that work, which is totally independent from ours, are 
in good agreement with ours.



\bibliographystyle{apsrev4-1}
\bibliography{main}

\end{document}